\documentclass[aps,prb,twocolumn,superscriptaddress]{revtex4}
\usepackage{color}
\usepackage{graphicx}

\begin{document}
\title{Nonlocal resistance and its fluctuations in microstructures of band-inverted HgTe/(Hg,Cd)Te quantum wells}

\author{G. Grabecki}
\email{grabec@ifpan.edu.pl}
\affiliation{Institute of Physics, Polish Academy of Sciences,  PL-02 668 Warszawa, Poland}
\affiliation{Department of Mathematics and Natural Sciences, College of
Sciences,
Cardinal Wyszy\'nski University, PL 01-938 Warszawa, Poland}
\author{J. Wr\'obel}
\affiliation{Institute of Physics, Polish Academy of Sciences,  PL-02 668 Warszawa, Poland}
\affiliation{Faculty of Mathematics and Natural Sciences, Rzesz\'ow
University,  PL-35 959 Rzesz\'ow, Poland}
\author{M. Czapkiewicz}
\affiliation{Institute of Physics, Polish Academy of Sciences,  PL-02 668 Warszawa, Poland}
\author{{\L}. Cywi\'nski}
\email{lcyw@ifpan.edu.pl}
\affiliation{Institute of Physics, Polish Academy of Sciences,  PL-02 668 Warszawa, Poland}
\author{S. Giera{\l}towska}
\affiliation{Institute of Physics, Polish Academy of Sciences,  PL-02 668 Warszawa, Poland}
\author{E. Guziewicz}
\affiliation{Institute of Physics, Polish Academy of Sciences,  PL-02 668 Warszawa, Poland}
\author{M. Zholudev}
\affiliation{L2C, UMR N$^o$5221 CNRS, Universit\'e Montpellier 2, GIS-TERALAB,
F-34095 Montpellier, France}
\affiliation{Institute for Physics of Microstructures, Russian Academy of
Sciences,  GSP-105, Nizhny Novgorod, 603950, Russia}
\author{V. Gavrilenko}
\affiliation{Institute for Physics of Microstructures, Russian Academy of
Sciences,  GSP-105, Nizhny Novgorod, 603950, Russia}
\author{N. N. Mikhailov}
\affiliation{Institute of Semiconductor Physics, Siberian Branch, Russian
Academy of Sciences, Novosibirsk, 630090, Russia}
\author{S. A. Dvoretski}
\affiliation{Institute of Semiconductor Physics, Siberian Branch, Russian
Academy of Sciences, Novosibirsk, 630090, Russia}
\author{F. Teppe}
\affiliation{L2C, UMR N$^o$5221 CNRS, Universit\'e Montpellier 2, GIS-TERALAB,
F-34095 Montpellier, France}
\author{W. Knap}
\affiliation{L2C, UMR N$^o$5221 CNRS, Universit\'e Montpellier 2, GIS-TERALAB,
F-34095 Montpellier, France}
\affiliation{Institute of High Pressure Physics, Polish Academy of Sciences, PL 01-142 Warszawa, Poland}
\author{T. Dietl}
\affiliation{Institute of Physics, Polish Academy of Sciences,  PL-02 668 Warszawa, Poland}
\affiliation{Institute of Theoretical Physics, Faculty of Physics, University
of Warsaw,  PL-00 681 Warszawa, Poland}
\affiliation{WPI-Advanced Institute for Materials Research (WPI-AIMR), Tohoku
University, Sendai 980-8577, Japan}

\date{\today}

\begin{abstract}
We investigate experimentally transport in gated microsctructures containing a band-inverted HgTe/Hg$_{0.3}$Cd$_{0.7}$Te quantum well. Measurements of nonlocal resistances using many contacts prove that in the depletion regime the current is carried by the edge channels, as expected for a two-dimensional topological insulator. However, high and non-quantized values of channel resistances show that the topological protection length ({\em{}i.e.} the distance on which the carriers in helical edge channels propagate without backscattering) is much shorter than the channel length, which is $\sim \! 100$ $\mu$m. The weak temperature dependence of the resistance and the presence of temperature dependent reproducible {\em quasi}-periodic resistance fluctuations can be qualitatively explained by the presence of charge puddles in the well, to which the electrons from the edge channels are tunnel-coupled. 
\end{abstract}

\pacs{72.25.Dc,73.23.Hk,73.61.Ga,73.63.Hs}

\maketitle

\section{Introduction}
The most spectacular property of two-dimensional (2D) topological insulators (TIs) is the presence of counter-propagating  edge channels of a specific spin (or pseudo-spin) structure,\cite{Kane2005,Bernevig2006}  which account for the persistence of conductivity even when the interior of the sample is  depleted of carriers, as demonstrated experimentally\cite{Koenig2007,Roth2009} for quantum wells (QWs) of HgTe with inverted band structure ({\em{}i.e.}~the thickness of which is above a critical value of $6.3$~nm). At a given energy these \emph{helical} electron states propagating along the channel in the two opposite directions form a Kramers pair\cite{Kane2005} and are, therefore, protected by the time-reversal symmetry (TRS) against \emph{elastic} back-scattering.\cite{Kane2005,Bernevig2006}

Reflectionless one-dimensional (1D) transport along edges should lead to a perfect quantization of conductance, that is, the emergence of the quantum spin Hall effect\cite{Bernevig2006} (QSHE) when the 2D sample is in the TI regime. Indeed, experiments with HgTe QWs of the appropriate thickness provided the evidence of highly conducting edge states in two-terminal\cite{Koenig2007} and also multi-terminal, non-local transport measurements.\cite{Roth2009}  Similar results were also obtained for microstructures of InAs/GaSb QWs.\cite{Knez2011,Suzuki_PRB13,Du_arXiv13}  
However, the values of conductance $G$, measured when the Fermi level $\epsilon_{\text{F}}$ was tuned by the gate to be inside the energy gap of the QW, were only approximately equal to the expected quantized value $G_0 \! = \! 2e^2/h$.
This is in contrast to the  the accuracy achieved in the case of the quantum Hall effect\cite{Koenig2007,Zholudev2012} (QHE) or the quantized anomalous Hall effect,\cite{Chang:2013_S} for which the topological protection is ensured by a large (possibly macroscopic) spatial separation of the \emph{chiral} channels from the two edges of the sample, and, for QHE, by the presence of the gap between Landau levels leading to exponential suppression of temperature dependence of deviations from perfect quantization. In QSHE only the elastic backscattering in the channels is forbidden, and perfect quantization is expected only at low (possibly extremely low) temperatures.
In fact, even at temperature $T \! \sim \! 1$~K, the conductance showed fluctuations as a function of the gate voltage $V_{\mathrm{g}}$.\cite{Koenig2007,Brune:2012_NP,Suzuki_PRB13}  Furthermore, $G$ values were close to $G_0$ in HgTe QWs only\cite{Koenig2007,Roth2009} for samples with sizes between $1\times 1$~$\mu$m$^2$ and $5\times 5$~$\mu$m$^2$, whereas for larger devices conductance strongly decreased, down to $10^{-4}$~$G_0$ in structures with the channel length of the order of 1~mm.\cite{Gusev2011} 

When only elastic scattering is considered, the presence of TRS breaking (by external magnetic field $B$ or polarization of magnetic impurities or nuclear spins in the sample) is necessary for any kind of disorder at the edge to lead to backscattering (for calculations at finite $B$ with specific models of disorder see {\em{}e.g.}~Refs.~\onlinecite{Tkachov_PRL10} and \onlinecite{Delplace2012}). 
In the absence of the $B$ field, there are many theoretical models showing that inelastic processes, either due to electron-electron,\cite{Kane2005,Xu2006,Wu_PRL06,Schmidt2012} electron-phonon,\cite{Budich2012} and electron-magnetic impurity\cite{Maciejko_PRL09,Tanaka2011,Lunde2012} scattering, can lead to a decrease of conductance of helical edge channels. However, the observed weak temperature dependence\cite{Nowack2012,Gusev_arXiv13} (see also below) of this conductance rules out all these mechanisms. The current-induced nuclear polarization is theoretically possible,\cite{Lunde2012,DelMaestro2012} leading to breaking of TRS and to backscattering {\it via} Rashba interaction, but neither signatures of such TRS breaking, nor nonlinear I-V characteristics,\cite{DelMaestro2012} have been observed.

All the above-mentioned mechanisms can be labeled as \emph{intrinsic} to the helical edge, {\em i.e.}~the relevant inelastic scattering processes involve the carriers propagating through the edge states. Interestingly, the most plausible model explaining the non-quantized resistance in the QSHE regime is of the \emph{extrinsic} character. Experimental data\cite{Roth2009,Konig_PRX2013} suggests that the edge states are coupled to charge puddles within the QW. 
Once an electron  finds  itself inside the puddle tunnel-coupled to the edge states, it can dwell there for a time longer than the inelastic scattering lifetime.\cite{Folk_PRL01}
If both the phase and the spin of the electron are lost during this dwell time, one can treat the puddle as an ``unintentional contact'' in which the chemical potentials for both spin directions are equalized.\cite{Roth2009,Konig_PRX2013} Deviations from perfect conductance observed in rather short channels could be explained by assuming the existence of a single puddle of this kind located close to the edge of the sample.\cite{Roth2009,Konig_PRX2013} 
Recently, a detailed theory of enhancement of inelastic backscattering due to dwelling of carriers in the puddles was given for the case of $T \! < \! \delta$ (where $\delta$ is the mean spacing of levels in the puddle), and the resistance of a long helical channel tunnel-coupled to multiple puddles was predicted to be $R \sim L n_{p}(T/\delta)^{3}$, where $L$ is the length of the channel and  $n_{p}$ is the density of puddles.

Let us note that the existence of such unintentional quantum dots is natural\cite{Vayrynen:2013_PRL} for a modulation-doped system with a small bandgap\cite{Novik_PRB05,Beugeling:2012_PRB} ($E_{g} \!\sim \! 10$ meV for HgTe/HgCdTe QWs and $E_{g} \! \leq \! 4$ meV for InAs/GaSb structures\cite{Knez2011,Du_arXiv13,Suzuki_PRB13}), in which the Coulomb disorder from ionized dopants leads to large (relative to $E_{g}$) spatial fluctuations of band edge energies. While signatures of presence of a single charge puddle coupled to a helical edge were seen in experiments on HgTe/HgCdTe wells,\cite{Roth2009} there are less data supporting the existence of charge puddles in InAs/GaSb systems supporting the TI state. However, the recent increase\cite{Du_arXiv13} in the accuracy of conductance quantization obtained due to introduction of Si doping at the interface (with Si acting as a donor in InAs and an acceptor in GaSb) is consistent with the impurities acting as localization centers depleting carriers from the larger puddles. It should also be noted that an additional source of inhomogeneity in 2D TI samples are fluctuations of bandgap size due to inhomogeneities in the QW width, the presence of which was suggested in the HgTe/HgCdTe system.\cite{Konig_PRX2013} The main observation is that the situation in these very narrow bandgap material is analogous to the situation in graphene, in which strong carrier density inhomogeneity well-known to exist in the gapless case\cite{DasSarma_RMP11,Martin_NP08} also persists for finite but small bandgap in gapped bilayer graphene.\cite{Rossi_PRL11}

In this work we present resistance and magnetoresistance  data for gated multi-probe Hall micro-bridges of HgTe/Hg$_{0.3}$Cd$_{0.7}$Te QWs with the edge channel length of the order of 100~$\mu$m, that is, in-between the length scales explored previously.\cite{Koenig2007,Roth2009,Gusev2011} Our non-local resistance data obtained for various combinations of contacts provide a direct proof that the current flows along edges when the QW interior is depleted from carriers (or, more precisely, made insulating by the gate voltage), 
corroborating the results of recent experiments aimed at imaging the edge states with various probes.\cite{Konig_PRX2013,Nowack2012,Ma2012}
However, the resistance magnitudes are systematically higher than the quantized values expected for reflectionless edge channels and, moreover, they are weakly temperature dependent. 
These results can be explained by the coupling of helical edge channel to charge puddles resulting from the potential fluctuations, which are only weakly screened by the metal gate
located $160$ nm above the QW. The spacing between the puddles along the channel that we can infer from our data is $< \! 10$ $\mu$m. Furthermore, the presence of charge puddles can also explain reproducible resistance oscillations observed in  the depletion regime. These oscillation  exhibit visible periodicity, which can be explained by the  discreteness of energies required for  carrier addition/tunnelling to  particularly small puddles.

\section{Samples}
\begin{figure*}[tb]
\centering
\includegraphics{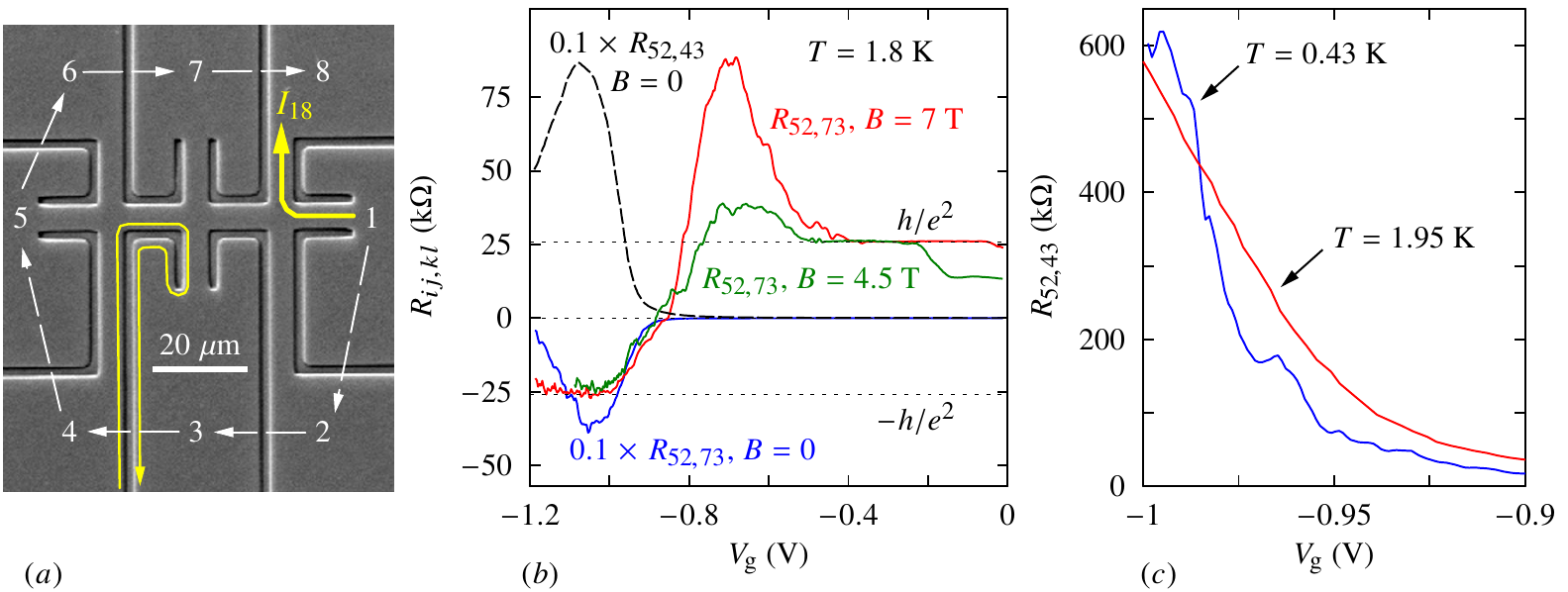}
\caption{(color online). (a) Scanning electron micrograph of HgTe/HgCdTe structure No.~1. Etched trenches define the Hall bar of a nominal channel width of $5$~${\mu}$m; the size of the image corresponds roughly to the top gate area. 
White arrows indicate a monotonous decrease of electrical potential of the contact probes, expected if the current flows only along the edges. Thin yellow arrow shows one of the edge channels. (b) Four-probe longitudinal and Hall resistances as a function of gate voltage for structure No.~1. (c) Local resistance in structure No.~1 for two temperatures.}
\label{fig:1}
\end{figure*}

Our samples are patterned from heterostructures grown by molecular beam epitaxy (MBE) on semi-insulating GaAs [013] substrates with a relaxed CdTe buffer.\cite{Dvoretsky2010} They consist of an 8~nm thick n-HgTe QW residing between Hg$_{0.3}$Cd$_{0.7}$Te barriers modulated in central parts by In dopants. Their characterization by the QHE was already reported.\cite{Zholudev2012} The microstructurization procedure has been performed by electron beam lithography with precautions taken not to overheat the structure during the resist annealing ($T \le 80^o$ C). The pattern, shown in Fig.~\ref{fig:1}(a), consists of $0.9$~${\mu}$m deep trenches forming an eight probe Hall bar structure, obtained by wet chemical etching in the solution of $0.05$ \% Br in ethylene glycol. After etching, the whole structure was covered by $100$~nm thick layer of HfO$_2$/Al$_2$O$_3$ composite, grown by atomic layer deposition.\cite{Gieraltowska2011} Subsequently, $30$~nm thick gold film of an area $90 \times 90\,{\mu}$m$^2$ was deposited in the central part of the structure. Finally, the insulator layer on the large contact pads was mechanically punctured and indium contacts were attached by soldering. Two structures of differing distance between the probes $D$ and Hall bar width $W$ have been fabricated: for structure No.~1 we have $D \! \approx \! 15$ $\mu$m and $W\lesssim 5$~${\mu}$m, while for structure No.~2 we have $D \! \approx \! 6$ $\mu$m and $W\lesssim 2$~${\mu}$m.

In contrast to mesa-like samples used in the previous works,\cite{Koenig2007,Gusev2011} the edge channels for our layout are expected to form on both sides of the insulating lines, as shown schematically in Fig.~\ref{fig:1}(a). The total length of the channels between adjacent contacts (see Fig.~1(a)) is estimated to be $L \! \sim \! 100$~${\mu}$m, a value by a factor of $10$ larger and smaller than those specific to samples from Ref.~\onlinecite{Roth2009} and Ref.~\onlinecite{Gusev2011}, respectively. Resistance measurements have been performed by low frequency ac method ($f \lesssim 10$~Hz) in the temperature range between $ 0.3 $ and $ 4.2 $~K in a $^3$He system. Typically, the current excitation voltage has been kept below $ 100 $~$\mu$V, and the linearity of $I-V$ characteristics has been checked. Gate voltages up to $|V_{\text{g}}|= 1.2$~V have been employed, for which the gate leakage current is below our detection limit, $ 0.1 $~nA.

From the Hall and resistance data in low magnetic fields and with no gate voltage, the 2D electron concentration  and mobility  are $n_{\text{2D}}=2.5\times 10^{11}$~cm$^{-2}$, ${\mu}=1.0\times 10^{5}$~cm$^{2}$/Vs (structure No.~1) and  $n_{\text{2D}}=1.6\times 10^{11}$~cm$^{-2}$, ${\mu}=2.4\times 10^{4}$~cm$^{2}$/Vs (structure No.~2), corresponding to the mean free path $\ell = 1.0 $ and $0.12 $~$\mu$m, respectively. For comparison, for bulk HgTe, $\ell$ reaches 2~$\mu$m,\cite{Dubowski:1981_JPCS} whereas in (Hg,Mn)Te near the Dirac point, $\ell$ up to 13~$\mu$m was found.\cite{Sawicki:1983_Pr}

\section{Magnetotransport and local resistance}
For the two fabricated structures, we can achieve the depletion regime where the conductance type changes from $n$ to $p$, and the Fermi level is expected to pass through  the energy bandgap of the QW. This  is illustrated  in Fig.~\ref{fig:1}(b), where we show the measured resistances $R_{ij,kl}$ for current flowing between contacts $(ij)$, with the voltage drop measured between probes $(kl)$. The data are taken for structure No.~1, but the qualitative features discussed below, with the exception of the presence of QHE plateaus, were also observed in structure No.~2. The lack of QHE in structure 2 (with $W \! \lesssim \! 2$ $\mu$m) suggests an important role of disorder and associated charge inhomogeneities possibly coupling the opposite edges of this narrower structure.

In Fig.~\ref{fig:1}(b) we show the gate voltage dependence of the Hall resistance $R_{52,73}$ (measured for magnetic field $B \! = \! 0$, $4.5$, and $7$ T), and of the local resistance $R_{52,43}$ for structure No.~1. In the n-type regime (small magnitudes of $|V_{\text{g}}|$)), two or one  well developed QHE plateaus ($h/2e^2$ and $h/e^2$) have been observed at 4.5~T and 7~T, respectively, pointing to a sizable topological protection length of chiral edge channels. 
With a further increase of the electron depletion, the Hall resistance $R_{52,73}$ goes through a maximum and changes its sign for both values of $B$ at $V_{\text{g}} \! \approx \! -0.9$~V. The $V_{\text{g}}$ range of large longitudinal (local) resistance, $R_{52, 43}$ is associated with carrier depletion of the QW, or, more precisely, with the suppression of bulk conductivity.
Let us note that the maximum of the zero-field local resistance is at $V_{\text{g}} =-1.1$~V, while the Hall voltage changes sign at $V_{\text{g}} \! \approx \! -0.9$~V. This not surprising, since  the magnetic field-induced band evolution in HgTe QW leads to substantial quantitative modification of the size of the bandgap,\cite{Novik_PRB05,Beugeling:2012_PRB} so that the depletion of the QW interior occurs at different $V_{\text{g}}$ in zero and nonzero $B$ field.

The maximum resistance level, $R_{52,43}^{\text{max}}{\approx} 800$~k$\Omega $, is much higher than the quantized values observed in Ref.~\onlinecite{Roth2009}. We have checked  that similarly high values are observed for $R$ measured with other probes, {\em e.g.}, $R_{58,32}^{max}{\approx} 300$~k$\Omega $.  Therefore, we observe a behavior similar to that reported\cite{Gusev2011} in  structures with 1~mm-long edge channels: the local resistance in the depletion regime is much higher than the quantized value seen in small samples,\cite{Koenig2007,Roth2009} 
and it shows very weak temperature dependence, as illustrated in Fig.~\ref{fig:1}(c).
It should be noted that the ratio of typical magnitudes of resistances shown here and those from Ref.~\onlinecite{Gusev2011} is similar to the ratio of the relevant channel lengths ($\sim \! 10^{2}$ $\mu$m here and $\sim \! 10^{3}$ $\mu$m in Ref.~\onlinecite{Gusev2011}), suggesting that the long edge channels in these structures behave to a large degree as classical (\emph{i.e.}~incoherent) wires (see Section V for further discussion). 
However, let us note one notable difference between our results and those from Ref.~\onlinecite{Gusev2011}, which is the presence of reproducible fluctuations of the resistance visible at low $T$ in Fig.~\ref{fig:1}(c).

\begin{figure}[tb]
\centering
\includegraphics{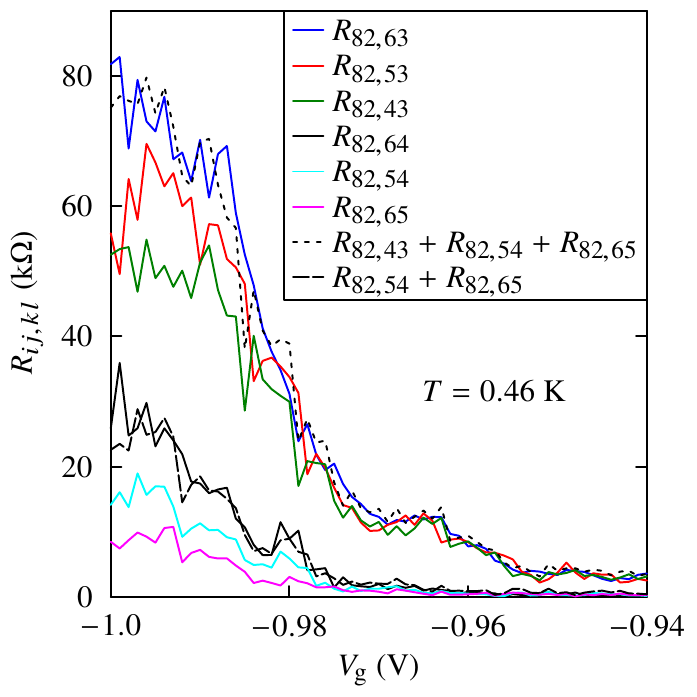}
\caption{(color online).  Nonlocal resistances $R_{ij,kl}$ at 0.46~K {\em vs.} gate voltage $V_{\mathrm{g}}$, measured for fixed current contacts $(ij)\! = \! (82)$ and for a series of the voltage probes $(kl)$ located around structure No.~1. The sign and the additivity of $R_{ij,kl}$ values demonstrate that the current flows only along edges in this range of gate voltages but the magnitudes of $R_{ij,kl}$ are higher than expected for relectionless channels. }
\label{fig:2}
\end{figure}

\begin{figure}[tb]
\centering
\includegraphics{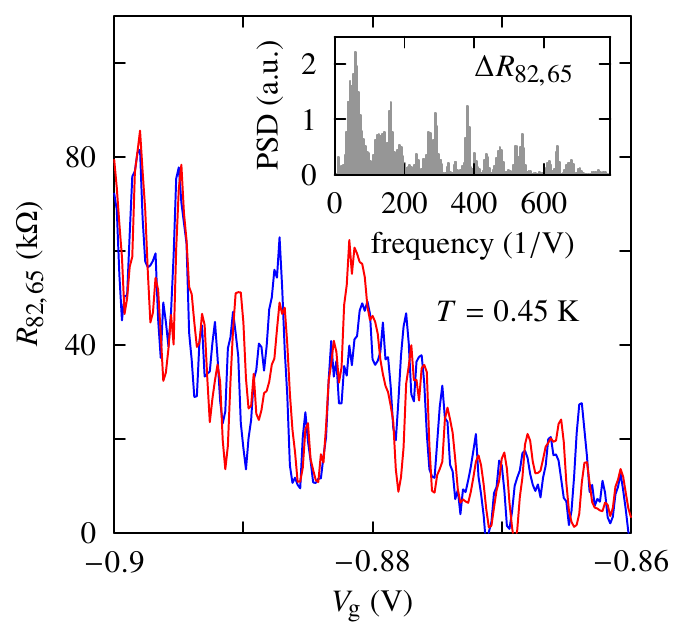}
\caption{(color online). Nonlocal resistance measured for opposite directions of gate voltage sweep showing the reproducible character of resistance fluctuations, and also their quasi-periodicity. Inset: spectral density of fluctuations showing the presence of multiple characteristic frequencies. The results are for sample No.~1, but were taken during a different cool-down cycle than the one presented in Fig.~\ref{fig:2}. The $V_{\text{g}}$ corresponding to QW depletion was observed to significantly vary from one cool-down to another. }
\label{fig:3}
\end{figure}

\section{Nonlocal resistance}
In order to demonstrate that in the $V_{\text{g}}$ range corresponding to the depletion conductance proceeds by edge channels, we have studied the nonlocal resistance $R_{\text{nl}}$.
For H-type structures, the contribution of carriers from the QW gives $R_{\text{nl}} \simeq R_{\text{sq}}\exp(-\pi D/W)$, where $R_{\text{sq}}$ is the QW resistivity; $D$ and $W$ are the bar length and width, respectively.
Dependencies of nonlocal resistances $R_{ij,kl}$ on  $V_{\text{g}}$ for $B \! = \! 0$ are presented in Fig.~\ref{fig:2} for structure No.~1. We have used contacts $(ij)\! = \!(82)$ as current leads and have measured voltage employing probes $(kl)$ along the structure perimeter.
The most important observation is {\em the same sign of the voltages} measured between the subsequent pairs of the probes. This indicates that the current flows in the same direction around the structure, {\em i.e.}, along the path $6-5-4-3$, and its direction is opposite to that of the current directly flowing between $8-2$.  In Fig.~\ref{fig:2} we show six curves that represent results for various pairs of the voltage probes located along the clockwise current path between the $(82)$ contacts.
Importantly, the data show that all subsequent nonlocal resistances are additive, {\em e.g.}, $R_{82,63}=R_{82,43}+R_{82,54}+R_{82,65}$, as  is illustrated by the dotted and dashed curves in Fig.~\ref{fig:2}.   
We have also checked that the ``sum rule'' of subsequent voltages holds when the temperature is increased up to 1.8~K.
An analogous behavior has been observed for non-local resistances measured for structure No.~2 at 1.8~K. There, the current has been fed through the leads $(ij) = (18)$. As in sample No.~1, in this narrower structure the voltage sign along the path $8-7-...-2-1$ does not change and corresponds to a current circulating around the QW edges between probes $8$ and $1$. Similarly to the case of the local resistance,  we have observed a rather weak $T$ dependence of $R_{ij,kl}$ between $0.4$ to $1.8$~K for both structures.

While some results of $R_{\text{nl}}$ published so far confirm the presence of the edge conduction in HgTe QWs, they were obtained only for selected pairs of current and voltage probes.\cite{Roth2009} Such selective measurements would prove the existence of the edge channel only if there are no other parallel conducting channels in the structure. Otherwise, a false nonlocal signal may occur, as it was recently demonstrated in Hall microstructures of PbTe.\cite{Kolwas2013}  Therefore, only a series of nonlocal voltage measurements, taken between consecutive contact probes distributed around the structure perimeter, confirms the edge conduction. Here we have shown that in the depletion regime, electric potential measured around the circumference of the structure is really compatible with the edge conduction picture.

For the reflectionless channels\cite{Roth2009} and the eight probe geometry in question, the quantized value of the nonlocal resistance measured with consecutive voltage probes along the $2$-$3$-...-$7$-$8$ path (with $(82)$ as current leads) should be equal to $h/4e^2\!=\! 6454$~$\Omega$. The values of $R_{82,43}$ and $R_{82,54}$ are much larger than this value, and $R_{82,65}$ is somewhat larger. Note that the latter observation \emph{does not} mean that the channel between contacts $5$ and $6$ is (nearly) reflectionless, since the value of nonlocal resistance depends on the resistances of all the other segments of the circuit, and once the quantization expected for perfect channels is broken (as is the case here), the value of $R_{\text{nl}} \! \approx \! h/4e^2$ loses any particular significance. We define $R_{nm}$ as the resistance of the clockwise channel between $m$ and $n$ contacts, {\em{}i.e.}~the ideal two-terminal resistance  for current flowing \emph{only} along $m$-$m$+$1$-...-$n$ path, see Fig.~\ref{fig:1}(a) for contact labels. For the case of $l$-$l$+$1$-...-$k$ path contained in the $j$-$j$+$1$-...-$i$ path ({\em{}i.e.}~for local and nonlocal resistances considered here) we have
\begin{equation}
R_{ij,kl} = \frac{R_{kl}}{R_{ij}} \frac{1}{R^{-1}_{ij} +  R^{-1}_{ji}} \,\, . \label{eq:R}  
\end{equation}
From this equation one can see that while the total values of $R_{ij,kl}$ depend on the resistances of all the segments of the channel encircling the QW, the ratios of these resistances taken for various $(kl)$ while keeping $(ij)$ fixed correspond to the ratios of respective $R_{kl}$ resistances. Using this observation, from Fig.~\ref{fig:2} we can read off the approximate ratios $R_{43} \! \approx \! 7 R_{65}$ and $R_{54} \! \approx \! 2R_{65}$. Since the channels connecting the consecutive contacts are of comparable length, this means that at the length scale of about $100$ $\mu$m the Ohmic scaling of the channel resistance with its length is not yet valid, i.e.~the contributions from scattering centers (the nature of which is discussed below) are not yet self-averaging. In the next Section we will discuss this observation further.

Finally, we note again the presence of resistance fluctuations at temperature $T \! < \! 1$ K, shown for the nonlocal signal in Figures \ref{fig:2} and \ref{fig:3}. As one can see in the inset of Fig.~\ref{fig:3}, these fluctuations contain a few components periodic in the gate voltage $V_{\text{g}}$.

\section{Discussion of the mechanism leading to large edge channel resistance}
From the results presented above we conclude that: (1) the current in the depletion regime flows along the edge of the sample, (2) the edge channel resistance is much higher than the value corresponding to helical channels free of backscattering, and it is very weakly (sublinearly) dependent on temperature in range $0.4$ K $< \! T \! < \! 1.8$ K, 
and (3) there are reproducible resistance fluctuations which become less pronounced as temperature is increased.
 If we discard the possibility that the \emph{trivial}, {\em{}i.e.}~non-helical, edge channels are formed in an inverted QW (which would be a truly unpleasant coincidence, but not impossible in principle, especially in the light of the uncovered presence of band bending\cite{Ma2012} at the sample boundaries in such QWs), the measured values of resistance imply an inelastic scattering mechanism for the edge carriers being quite efficient for channel lengths of $\gtrsim \! 10$ $\mu$m. 
In fact, even if the edge channel was non-helical, the large but finite value of resistance 
would be consistent with the well-established physics of Anderson localization in 1D\cite{Imry} only in the presence of rather strong mechanism of phase-breaking. 
The strong localization is present only when the localization length $l_{\text{loc}}$ is smaller than the phase-breaking length $l_{\phi}$. A large ($R \! \gg \! h/e^{2}$) value of resistance and $R \! \sim \! L$ scaling (suggested by comparison of our results to those from Ref.~\onlinecite{Gusev2011}) are consistent with dominantly inelastic character of scattering. In such a case $l_{\phi}$ is of the same order of magnitude as the scattering length, and since the localization in 1D is due to interference of electron trajectories involving multiple scatterings, the localization is preempted by phase breaking.
However, if the edge is of helical character, the weak $T$-dependence of the resistance might be considered puzzling in the light of super-linear dependence predicted by many theoretical models.\cite{Kane2005,Xu2006,Tanaka2011,Schmidt2012} The presence of reproducible fluctuations of both local and nonlocal resistance, see Figs.~\ref{fig:2} and \ref{fig:3}, also demands then an explanation which is consistent with phase-breaking length being much shorter than the channel length.

A natural explanation for these observations is provided by the assumption of existence of charge puddles coupled to the edge states.\cite{Roth2009,Konig_PRX2013,Vayrynen:2013_PRL}
When the inelastic scattering time within the puddle is at least comparable to the mean dwell time inside of the dot, the carrier has randomized phase when it tunnels back into the edge channel. The spin is also expected to be randomized, since in structures of narrow bandgap semiconductors such as HgTe/HgCdTe and InAs/GaSb the spin-orbit induced spin splittings of lowest energy carrier subbands are significant (for example, Rashba splitting can reach $30$ meV in gated HgTe/HgCdTe quantum well\cite{Gui_PRB04}). 
The  weak $T$ dependence of resistance can then be naturally explained: once the puddles act as (nearly) perfect phase- and spin-randomizing contacts, a further increase of $T$ does not affect the resistance in a visible manner.
The condition for reaching this regime is $k_{\text{B}}T \! > \! \delta$, for which the theory from Ref.~\onlinecite{Vayrynen:2013_PRL} predicts a saturation of $T$-dependence of resistance.\cite{Glazman_unpublished}

When both the phase and the spin is randomized during the carrier dwell time, each puddle acts as an unintentional contact\cite{Roth2009} allowing for backscattering and increasing the helical channel resistance by $h/e^{2}$. The measured resistance is then given by Eq.~(\ref{eq:R}), in which $R_{kl}  \! =\! (n_{kl}+1)h/e^2$ where $n_{kl}$ is the number of dephasing sites between the $(kl)$ contacts weighted by the probability of tunneling into them, so that $n_{kl}$ decreases with the channel length but remains non-zero and fluctuates with the Fermi energy and magnetic field even for short channels.
As we have discussed in the previous section, the values of $R_{kl}$ for various pairs of consecutive contacts have a wide distribution, \emph{e.g.}~$n_{43} \! \approx \! 6(n_{65}+1) \! \approx \! 3 (n_{54}+1)$. These ratios cannot be explained by ratios of channel lengths between adjacent contacts, which are all between $80$ and $120$ $\mu$m. Also, from Eq.~(\ref{eq:R}) we see that $R_{kl} \! > \! R_{ij,kl}$, so that the local resistance measurement shown in Fig.~1(b) gives $n_{43} \! > \! 28$. Thus, the spread of $n_{kl}$ values cannot be explained by simple statistical fluctuations $\sim \! \sqrt{n_{kl}}$ expected for Poisson statistics of $n_{kl}$ in the case of a fixed rate of strong coupling to a puddle per unit length of the channel. Together with the fact that the ratio $R_{52,43}/R_{82,43}$ shows that $R_{28}/R_{25}$ is only about $1/15$ instead of $\approx \! 2/5$ expected from geometry of the structure, the above observations suggest a presence of inhomogeneity of puddle density over the area of the sample shown in Fig.~1(a). Such an inhomogeneity makes it harder to extract precise value of channel resistances, such as $R_{43}$, from the data presented here. We can only say that the lower bound for $n_{43}$ given above means that the distance between puddles acting as full dephasing centers ({\em{}i.e.}~the topological protection length) can be as short as a few $\mu$m. 
 
The conjectured inhomogeneity of puddle density could be due to fluctuations of quantum well width $d$, which should lead to large relative fluctuations of the magnitude of the energy gap $E_{g}$ of the well - it should be kept in mind that $E_{g} \! = \! 0$ for $d\! = \! 6.3$ nm, while for $d\! = \! 8$ nm we should have $E_{g} \! \approx \! 10$ meV. Structural defects of planar size of $\sim\! 1 $ $\mu$m were in fact observed by AFM in other structures of HgTe/HgCdTe, and their relation to local modifications of quantum well bandstructure was suggested.\cite{Konig_PRX2013} According to estimates from Ref.~\onlinecite{Vayrynen:2013_PRL}, the typical puddle size is $\sim \! a_{\text{B}}$ with the effective Bohr radius $a_{\text{B}} \! \sim \! 1/E_{g}$. The presence of micron-sized areas of quantum well with $d$ closer to $6$ nm than to the nominal value of $8$ nm could then cause the appearance of rather large puddles (possibly merging into a micron-sized quasi-bulk region) with level spacing $\delta$ an order of magnitude smaller than $\sim \! 1$ meV estimated for $E_{g} \! = \! 10$ meV in Ref.~\onlinecite{Vayrynen:2013_PRL}. In such a case the temperature (in)dependence of our data would be qualitatively consistent with the microscopic theory of backscattering due to presence of charge puddles.\cite{Glazman_unpublished} Furthermore, the \emph{extrinsic} (not following from self-consistency of puddle filling and the resulting screening of external potential fluctuation in the presence of uniform density of ionized remote impurities)  origin of the particularly large puddles could help reconcile the large values of channel resistances with the lack of bulk conductivity due to merging of the puddles. If the regions of smaller $E_{g}$ containing more puddles (or possibly each one of them being simply uniformly filled with carriers) were rare enough to be uncoupled, and if they did not span the whole area between the edge channels from the opposite sides of the quantum well, their presence would not affect our observation of current flowing along the edges of the structure, with negligible contribution of transport through the bulk.

\begin{figure}[tb]
\centering
\includegraphics{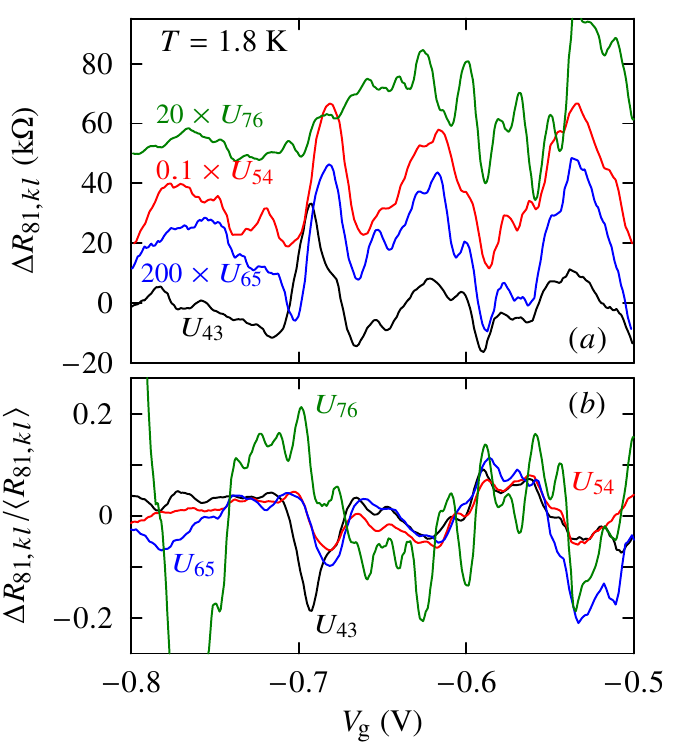}
\caption{(color online). (a) Nonlocal resistance fluctuations for four pairs of voltage probe contacts measured in structure No.~2, offset one with respect to another for clarity. (b) The same normalized by the smoothened value of the respective nonlocal resistance.  The relative fluctuations of $R_{81,76}$ are clearly distinct from the relative fluctuations of the four other resistances. }
\label{fig:4}
\end{figure}

Let us also note that while in Ref.~\onlinecite{Vayrynen:2013_PRL} predictions are given for both the mesoscopic regime (of a few puddles coupled to a channel) and the self-averaging regime of large $L$, the size of our structures seems to put them in an intermediate regime of moderate channel lengths. Multiple puddles contribute then to backscattering, but the fluctuations in resistance as a function of $V_{\text{g}}$, whether due to energy dependence of tunnel couplings, the discrete energy levels in the puddles, or the Coulomb blockade, are still visible. In fact, the measured resistance fluctuations are clearly showing signs of periodicity, as shown in Fig.~\ref{fig:3}. Furthermore, it is possible to identify the spatial location of sources of particularly prominent fluctuations.
In Fig.~\ref{fig:4}(a) we present the fluctuating parts of $R_{\text{nl}}(V_{\text{g}})$ in sample No.~2 for four pairs of voltage probes, with $(81)$ as the current-carrying contacts. For voltage probes $(65)$, $(54)$ and $(43)$ the fluctuations are very similar, with a quasi-periodic component (three peaks clearly visible in Fig.~\ref{fig:3}(a), the largest of which is at $V_{\text{g}} \! \approx \! -0.7$ V), 
while for the $(76)$ pair the fluctuation pattern is distinct. This strongly suggests that the source of resistance fluctuations is the \emph{global} modulation of the current flowing along the edge, caused by a localized $V_{\text{g}}$-dependent resistance source.
If we assume that the strongest modulation of $\Delta R$ is caused by a periodic in $V_{\text{g}}$ variation of the coupling to a single charge puddle (caused possibly by a Coulomb blockade oscillation of the tunneling into this puddle), then the results are consistent with this dominant unintentional contact being located between probes $7$ and $6$, {\em i.e.}, there are gate-induced fluctuations $\delta R_{76}$ of $R_{76}$ resistance, which of course lead to the fluctuation $\delta R_{81}$ of the same magnitude.
Then, for the three pairs of probes (assuming for simplicity $R_{81} \! \gg \! R_{18}$, as is expected from the lengths of the respective current paths) we have for the strongest fluctuation $\delta R_{81,kl} \approx -\delta R_{81} R_{kl} R_{18}/{R_{81}^2}  \propto  R_{81,kl}$, while for the fourth pair we have $\delta R_{81,76} \approx -\delta R_{81}  R_{76}R_{18}/{R_{81}^2}  + \delta R_{76} R_{18}/R_{81}$, which is \emph{not} proportional to $R_{81,76}$. The measured relative fluctuations $\Delta R/R$ shown in Fig.~\ref{fig:4}(b) support this physical picture. 

\section{Conclusions}
The studies of nonlocal resistance in band-inverted HgTe/HgCdTe QWs show that when the bulk is depleted of mobile carriers, the current is indeed flowing around the sample edge. However, the channel length for which the backscattering is absent is of the order of a few $\mu$m, 
and the observed resistances are much larger than the quantized values predicted for helical channels unaffected by inelastic scattering. Weak temperature dependence of the edge resistance (for $0.4$ K $< \! T \! <$ $1.8$ K), together with the presence of reproducible resistance fluctuations having strong periodic components, can be explained by the presence of multiple disorder-induced charge puddles, which are tunnel-coupled to the edge states. The results support an emerging picture\cite{Roth2009,Konig_PRX2013,Vayrynen:2013_PRL} of very narrow bandgap two-dimensional topological insulators as having a very inhomogeneous energy landscape. 
Furthermore, both the observed wide spread of values of resistances for edge channels of comparable length, and the weak temperature dependence of these resistances, suggest the presence of fluctuations of the width or composition of the quantum wells, the length-scale of which exceeds the length-scale of ``intrinsic'' potential fluctuations caused by  the presence of remote ionized impurities.
The main conclusion is that in order to maintain a dissipationless edge transport by helical channels at distances larger than a few $\mu$m, the bulk of the sample not only needs to be insulating, but it also needs to be \emph{locally} insulating, {\em i.e.}~there should be no unintentional charge puddles close to the edge of the sample. 

\section*{Acknowledgements} 
We thank P.~Brouwer for a helpful discussion, and L.~I.~Glazman for his comments and suggestions. This work was supported by by National Science Centre (Poland) under Grant  DEC-2012/06/A/ST3/00247, by  Regional Development Program (Poland),
Grant WND-RPPK 01.03.00-18-053/12, and by FunDMS Advanced Grant (No. 227690) of European Research Council within ``Ideas'' 7th Framework Programme of European Commission.

\end{document}